\documentclass[review, 12pt, sort&compress]{elsarticle}

\usepackage{epsfig}
\usepackage{amssymb}
\usepackage{lineno}

\biboptions{square, numbers}

\journal{Chemical Engineering Science}

\newcommand{\abs}[1]{\ensuremath{\mathnormal{\left| {#1} \right|}}}
\newcommand{\absnum}{\ensuremath{\mathnormal{N}}}
   \newcommand{\absnumvec}{\ensuremath{\mathbf{N}}}
\newcommand{\adsorption}{\ensuremath{\mathnormal{\Gamma}}}
\newcommand{\area}{\ensuremath{\mathnormal{F}}}
\newcommand{\chempot}{\ensuremath{\mathnormal{\mu}}}
   \newcommand{\chempotid}{\ensuremath{\mathnormal{\chempot_\mathrm{id}}}}
   \newcommand{\chempotres}{\ensuremath{\mathnormal{\chempot_\mathrm{res}}}}
\newcommand{\compa}{\ensuremath{\mathnormal{i}}}

\newcommand{\coordx}{\ensuremath{\mathnormal{x}}}
\newcommand{\coordy}{\ensuremath{\mathnormal{y}}}
\newcommand{\coordz}{\ensuremath{\mathnormal{z}}}
\newcommand{\cutoff}{\ensuremath{\distance_\mathrm{c}}}
\newcommand{\density}{\ensuremath{\mathnormal{\rho}}}

\newcommand{\differential}{\ensuremath{\mathnormal{d}}}
\newcommand{\distance}{\ensuremath{\mathnormal{r}}}
\newcommand{\eer}{\ensuremath{\mathnormal{\eta}}}
\newcommand{\entropy}{\ensuremath{\mathnormal{S}}}
\newcommand{\force}{\ensuremath{\mathnormal{f}}}
\newcommand{\helmholtz}{\ensuremath{\mathnormal{A}}}
\newcommand{\kboltz}{\ensuremath{\mathnormal{k}}}

\newcommand{\length}{\ensuremath{\mathnormal{l}}}
\newcommand{\liq}[1]{\ensuremath\mathnormal{{{#1}'}}}
\newcommand{\LJenergy}{\ensuremath{\mathnormal{\epsilon}}}
\newcommand{\LJsize}{\ensuremath{\mathnormal{\sigma}}}

\newcommand{\molnum}{\ensuremath{\mathnormal{n}}}
\newcommand{\normal}[1]{\ensuremath{\mathnormal{#1}_\mathrm{\nu}}}

\newcommand{\planar}[1]{\ensuremath{\mathnormal{{#1}_\parallel}}}
\newcommand{\potential}{\ensuremath{\mathnormal{u}}}
\newcommand{\pressure}{\ensuremath{\mathnormal{p}}}
\newcommand{\qq}[1]{{\lq}#1{\rq}}
\newcommand{\radius}{\ensuremath{\mathnormal{R}}}
   \newcommand{\capillarity}{\ensuremath{\mathnormal{\radius_\kappa}}}
      \newcommand{\invcapillarity}{\ensuremath{\mathnormal{\radius^{-1}_\kappa}}}
   \newcommand{\equimolar}{\ensuremath{\mathnormal{\radius_\density}}}
   \newcommand{\laplace}{\ensuremath{\mathnormal{\radius}}}

\newcommand{\surfacetension}{\ensuremath{\mathnormal{\gamma}}}
   \newcommand{\equimolartension}{\ensuremath{\mathnormal{\surfacetension_\density}}}
\newcommand{\tangential}[1]{\ensuremath{\mathnormal{#1}_\mathrm{\tau}}}
\newcommand{\temperature}{\ensuremath{\mathnormal{T}}}
\newcommand{\thickness}{\ensuremath{\mathnormal{s}}}
\newcommand{\tolman}{\ensuremath{\mathnormal{\delta}}}
\newcommand{\vap}[1]{\ensuremath\mathnormal{{{#1}''}}}

\newcommand{\volume}{\ensuremath{\mathnormal{V}}}

\begin{document}
\begin{frontmatter}

\title{Molecular simulation of nano-dispersed fluid phases}
 
\author{Martin Horsch\corref{cor}}
\author{Hans Hasse}
\address{Lehrstuhl f\"ur Thermodynamik, Technische Universit\"at Kaiserslautern, Erwin-Schr\"odinger Str.\ 44, 67663 Kaiserslautern, Germany}
\cortext[cor]{Corresponding author. E-mail: martin.horsch@mv.uni-kl.de; phone: +49 631 205 3227; fax: +49 631 205 3835.}

\begin{abstract}
Fluid phase equilibria involving nano-dispersed phases, where at
least one of the coexisting phases is confined to
a small volume, are investigated by molecular dynamics
simulation.
Complementing previous studies on nanoscopic droplets, simulation
volumes containing a nanoscopic gas bubble surrounded by a subsaturated liquid
phase under tension, i.e.\ at negative pressure, are conducted in the
canonical ensemble. The boundary conditions are chosen such that the phase
equilibrium at the curved interface is thermodynamically stable.
Two distinct size-dependent effects are
found: Curvature induces a subsaturation of the
system, leading to a smaller liquid density.
For the gas in the centre of the bubble, the small diameter has an
additional obverse effect, increasing its density.
The curvature dependence of the surface tension is discussed by
evaluating average radial density profiles to obtain the excess
equimolar radius, which is found to be positive, corresponding to a
negative Tolman length. 
\end{abstract}


\end{frontmatter}

\linenumbers

\section{Introduction}
\label{sec:introduction}

\noindent
Dispersed phases are ubiquitous both in nature and technological
applications. Their character poses a particular challenge to
thermodynamic approaches which attempt to reduce the
complexity of a system to a few macroscopic degrees of freedom.
Even in the most bulk-like central region of a nanoscopic bubble or
droplet, thermodynamic properties may deviate substantially from the bulk phase
under corresponding conditions. Interfacial properties may dominate,
and the heterogeneity of the dispersion further complicates its
thermodynamic description.

Phenomenological thermodynamics was applied to fluid interfaces by
Gibbs \cite{Gibbs78a}, whose approach ultimately succeeded due to the rigour
with which it unifies the macroscopic and microscopic points of view. In
particular, it reduces the phase boundary, which is continuous on the
molecular level, to a strictly two-dimensional dividing surface separating
two bulk phases. The deviation between the actual system and the
theoretical system, consisting of the two bulk phases only, serves as a
definition of interfacial excess quantities to which phenomenological
thermodynamic reasoning can be applied.

This reduction facilitates discussing and analysing systems
which contain a nano-dispersed phase, but it does so at a prize. The task of
representing physically complex behaviour is shifted to the interfacial
excess quantities. Such quantities, and particularly the surface tension and the
adsorption, have to account for all the aspects which distinguish, for instance, the bulk
metal from a metal nanoparticle, or the bulk vapour from a gas bubble that
contains a few molecules only. This explains why such fundamental and
apparently simple issues such as the dependence of the surface tension of
small gas bubbles and liquid droplets on their radius are still not
fully settled, despite having been on the agenda of scientific discussions
for decades.

Furthermore, for the development of molecular equations of
state \cite{CGJR90, MWF96, GS01}, which mostly aim at
describing the bulk phases, it is important to understand
how precisely the intermolecular interactions affect the association
of molecules to small nanoclusters, since the underlying thermodynamic perturbation
theory \cite{Zwanzig54, Wertheim84a} is based on a statistical-mechanical
cluster expansion \cite{Mayer37}.
In addition, a reliable description of natural phenomena such as
atmospheric nucleation, as well as engineering problems such as nucleate pool
boiling, spray cooling, or nucleation in expanding gases as it is ubiquitous in turbines, can only be
obtained on the basis of quantitatively accurate models for the thermodynamic properties
of the respective dispersed fluid phases, i.e.\ nanoscopic gas bubbles and liquid droplets.
For such studies, both static and dynamic
properties have to be captured, concerning physical objects which can fluctuate
significantly in their size and shape or even disappear in the blink of an eye.

It is therefore attractive to apply molecular simulation to study these problems,
supplementing experimental results where they are available, and replacing
them where suitable experiments have not yet been devised. Molecular dynamics (MD) simulation
is capable of elucidating the properties of nano-dispersed phases in equilibrium as
well as dynamic phenomena including nucleation, aggregation, coalescence, growth,
wetting, and drying, among many others, at molecular resolution.
Even complex scenarios, such as gold clusters with an organic protection layer,
are well accessible to MD simulation \cite{SPV07}. In a simulation, boundary conditions can be imposed
which would be hard or impossible to guarantee in an experimental setting. For instance,
transport processes can be sampled in a well-defined steady state by non-equilibrium MD
simulation, including the coupled heat and mass transfer occurring at interfaces \cite{HI96}
and during nucleation in a supersaturated vapour \cite{HV09b}. The critical nucleus of
a nucleation process, which corresponds to a free energy maximum and is therefore
thermodynamically unstable, can be investigated in detail by equilibrium
simulation of a small system in the canonical ensemble \cite{NJV09}.

As a massively-parallel high performance computing application, MD simulation scales well both in
theory and in practice. Up to trillions of interaction sites can be simulated \cite{EHBBHHKVHHBGNBB13}, so
that a single modelling approach can be employed from the nanometre up to the
micrometre length scale. As such, molecular simulation is a useful tool for
investigating the size dependence of interfacial effects. MD simulations of
the surface tension of curved vapour-liquid interfaces, comparing it with that of the planar
phase boundary, were already conducted in the 1970s \cite{RB77}. Many of the subsequent
contributions to this problem, in particular more recently, have been guided by the analysis
of molecular simulation results \cite{NJV09, SVWZB09, VHH09, BDOVB10, SMMMJ10, DB11b, TB11, HHSAEVMJ12, MJ12}.

The present work illustrates the contribution that molecular modelling and simulation
can make to the discussion of nano-dispersed phases, with a focus on MD simulation
of a gas bubble in equilibrium with a liquid at negative pressure. 
This case is both of fundamental scientific interest and technically
important, e.g.\ for cavitation.
In Section \ref{sec:theory}, a brief survey is given on the relevant aspects of the
theory of vapour-liquid interfaces, including the dependence of the surface tension
on curvature and its relation to the excess equimolar radius.
Section \ref{sec:method} introduces the employed molecular simulation methods.
Simulation results, consistently finding the excess equimolar radius to be positive,
are presented in Section \ref{sec:results}. A possible interpretation of the present
results is suggested in Section \ref{sec:discussion}, relating it to previous
work and leading to the conclusion which is given in Section \ref{sec:conclusion}.

\section{Thermodynamics of dispersed phases}
\label{sec:theory}

\subsection{Vapour-liquid surface tension}
\label{subsec:theory-a}

\noindent
The tension of a planar fluid interface can be defined in different
ways, following a thermodynamic or a mechanical approach.
Thermodynamically, the surface tension $\surfacetension$ can be
expressed by the partial derivative of the free
energy $\helmholtz$ over the surface area $\area$
at constant number of molecules $\absnumvec$ (of all components),
volume $\volume$, and temperature $\temperature$:
\begin{equation}
   \surfacetension = \left(\frac{\partial\helmholtz}{\partial\area}\right)_{\absnumvec, \volume, \temperature}.
\label{eqn:surfacetension-thermo}
\end{equation}
The surface free energy can then be obtained by integration
\begin{equation}
   \helmholtz_\area = \int_0^\area \surfacetension \, \differential\area,
\label{eqn:surface-free-energy}
\end{equation}
over a process during which the interface is created.

By molecular simulation, the thermodynamic surface tension can be
computed from the test area method \cite{DeMiguel08}, while grand
canonical Monte Carlo simulation can be employed to obtain
$\helmholtz_\area$ from the excess Landau free energy corresponding
to the respective density \cite{Binder82, SVWZB09}.

Neg\-lecting size effects on $\surfacetension$, the surface
free energy can be approximated by $\helmholtz_\area \approx
\surfacetension\area$. While such a simplification is
justified for macroscopic systems, it may violate the thermodynamics
of small systems \cite{Hill64}, where, in general,
significant finite size effects can be present even for planar phase
boundaries \cite{GF90, WLHH13}. 

For a mechanical definition, the surface tension is treated as causing
a force $\tangential\force$ acting in tangential direction (with
respect to the interface), i.e.\ a tendency of the interface to
contract. The mechanical surface tension
\begin{equation}
   \surfacetension = \frac{\tangential\force}{\length}
\label{eqn:surfacetension-mech}
\end{equation}
relates the magnitude of this force to the length of the contact line
$\length$ between the interface and the surface of another mechanical object, e.g.\ a
confining wall, on which the force $\tangential\force$ acts.

In a cuboid box with the extension
$\volume = \length_\coordx \times \length_\coordy \times \length_\coordz$,
which contains a planar interface normal to the $\coordz$ axis, the
interface and the two faces of the box which are normal to the
$\coordx$ axis have contact lines with an elongation of $\length_\coordy$,
cf.\ Fig.\ \ref{fig:kasten}.
Each of these faces (normal to $\coordx$) has an area of $\area_{\coordy\coordz} = \length_\coordy \times \length_\coordz$.
The tangential force $\tangential\force = \force_\coordx = \surfacetension\length_\coordy$
thus constitutes a negative (contracting) contribution to the pressure, acting in
tangential direction, i.e.\ in $\coordx$-direction here.

The surface tension can thus be obtained from the
deviation between the tangential and normal eigenvalues $\tangential\pressure$
and $\normal\pressure$ of the pressure tensor:
\begin{equation}
   \tangential\pressure - \normal\pressure = - \frac{\surfacetension\length_\coordy}{\area_{\coordy\coordz}} = - \frac{\surfacetension}{\length_\coordz}.
\label{eqn:difpressure}
\end{equation}
In the example discussed above, the tangential pressure
$\tangential\pressure = \pressure_\coordx = \pressure_\coordy$ acts in the
$\coordx$- and $\coordy$-directions parallel to the interface, while the
normal pressure acts in $\coordz$-direction perpendicular to the interface.
It is well known that for planar fluid phase boundaries, the thermodynamic and
mechanical definitions of $\surfacetension$ coincide \cite{SM91}.
In molecular simulation, where the pressure tensor is computed from
the virial, an approach referred to as the virial route relies on Eq.\ (\ref{eqn:difpressure})
to obtain the surface tension \cite{WTRH83, VKFH06}.

\subsection{Curved vapour-liquid interfaces}
\label{subsec:theory-bd}

\noindent
At the curved interface of a bubble or a droplet, the mechanical equilibrium
condition is characterized by the La\-place equation
\begin{equation}
   \Delta\pressure = \liq\pressure - \vap\pressure = \frac{2\surfacetension}{\laplace},
\label{eqn:laplace}
\end{equation}
where $\liq\pressure$ and $\vap\pressure$ denote the pressure in the liquid
and the vapour phase, respectively. The radius $\laplace$ for which this relation holds
is called the Laplace radius or the radius of the surface of tension.
The interface tends to contract, compressing the dispersed phase which is situated inside,
and the surface tension $\surfacetension$ couples this
compressing effect with its cause, the curvature of the interface.
By convention, the radius $\laplace$ is positive
in case of a droplet (with $\liq\pressure > \vap\pressure$) and negative in
case of a bubble (with $\liq\pressure < \vap\pressure$).

It is worth recalling that within the thermodynamic approach
of Gibbs \cite{Gibbs78a}, the position of the formal dividing surface is
arbitrary at first. Thus, a further condition, such as Eq.\ (\ref{eqn:laplace}),
is needed to define a radius. The values of $\liq\pressure$ and
$\vap\pressure$ do not necessarily agree with the actual mechanical pressures on
the two sides of the interface. They are obtained
by combining the mechanical equilibrium condition, Eq.\ (\ref{eqn:laplace}),
with the chemical and thermal equilibrium conditions,
i.e.\ equal chemical potential $\liq{\chempot_\compa} = \vap{\chempot_\compa}$ for all components $\compa$
and equal temperature $\liq\temperature = \vap\temperature$.
The relation between the values of $\chempot_\compa$, $\pressure$, and $\temperature$
is given by the equation of state for the bulk phases.

For the case of a pure fluid below the critical temperature, a
$\chempot - \pressure$ diagram \cite{Debenedetti96} visualizes the impact of
curvature, by means of a vapour-liquid equilibrium condition with a pressure
difference between both phases, as expressed by Eq.\
(\ref{eqn:laplace}), on other thermodynamic properties such as the
density of the coexisting fluid phases and the chemical potential,
cf.\ Fig.\ \ref{fig:mu-p-diagram}.
The residual chemical potential $\chempotres$ is defined by the deviation of the
chemical potential $\chempot$ from its ideal temperature-dependent (i.e.\
density-independent) contribution $\chempotid$, reduced by temperature \cite{VH02}
\begin{equation}
   \chempotres(\density, \temperature) = \frac{\chempot(\density, \temperature) - \chempotid(\temperature)}{\temperature}.
\label{eqn:chempotres}
\end{equation}
At low densities it can be approximated by $\chempotres \approx
\ln\density$, so that the vapour parts of the three isotherms shown
in Fig.\ \ref{fig:mu-p-diagram} coincide roughly.
Its derivative with respect to pressure at constant temperature is
given by
\begin{equation}
   \left(\frac{\chempotres}{\pressure}\right)_\temperature = \frac{1}{\density\temperature}.
\end{equation}
Hence, proceeding (at increasing $\rho$) from stable vapour to
metastable vapour, to the unstable part of the isotherm, 
the metastable and finally the stable liquid, the slope of the curves
in the $\chempot - \pressure$ diagram decreases successively. In
Fig.\ 1 it can be seen how $\Delta\pressure = \liq\pressure - \vap\pressure > 0$,
corresponding to a droplet, induces a vapour-liquid equilibrium at a
supersaturated chemical potential with $\chempot > \chempot_\mathrm{sat}$,
where $\chempot_\mathrm{sat}$ is the chemical potential for the equilibrium
at a planar interface.
Obversely, in case of a bubble, the pressure is higher in the gas phase, i.e.\
$\Delta\pressure < 0$, so that the coexisting phases become
subsaturated ($\chempot < \chempot_\mathrm{sat}$).

The surface tension is then the differential
excess free energy (per surface area \area), so that the free energy of the whole system,
including the interface, is defined by
\begin{equation}
   \differential\helmholtz = \surfacetension \, \differential\area
                             - \entropy \, \differential\temperature
                             - \liq\pressure \, \differential\liq\volume
                             - \vap\pressure \, \differential\vap\volume
                             + \sum_\compa \chempot_\compa \, \differential\molnum_\compa.
\end{equation}
Therein, the entropy $\entropy$ also contains an interfacial excess term (which is
not relevant to the present discussion). The volume associated with the
interface, however, is zero, since the Gibbs dividing surface is
thought to be two-dimensio\-nal, so that the total volume $\volume = \liq\volume + \vap\volume$
is the sum of the liquid and vapour volumes.

While the thermodynamic and the mechanical approa\-ches to defining the
surface tension, see Eqs.\ (\ref{eqn:surfacetension-mech}) and (\ref{eqn:surfacetension-thermo}),
respectively, are strictly equivalent for planar fluid interfaces, cf.\ Section \ref{subsec:theory-a}, this
is not the case for solid systems, where the pressure tensor in the bulk
is not necessarily isotropic \cite{RTS10}. Also for nano-dispersed fluid phases, where an isotropic bulk-like
region may be completely absent, thermodynamic and mechanical definitions
of $\surfacetension$ deviate from each other \cite{HHSAEVMJ12, MJ12}:
Mechanical approaches following the virial route have found the
surface tension of nanodroplets to be significantly smaller than that of the planar
vapour-liquid interface \cite{TGWCR84, VKFH06}, whereas the thermo\-dynamic routes, i.e.\
the test area me\-thod \cite{SMMMJ10} and grand canonical Monte Carlo simulation \cite{BDOVB10},
do not confirm this and find such an effect to be much weaker or even of opposite sign.

An explanation of this disagreement between mechanical and thermodynamic expressions for
the surface tension is possibly to be found in the observation of Percus et al.\ \cite{PPG95}
that in general, the Landau free energy deviates from the volume integral over the
local pressure for inhomogeneous fluid systems. In any case, it is clear that the
quantity which is relevant to the Gibbs approach is the thermodynamic surface tension
and not the mechanical one.

Properties related to the smallest clusters, i.e.\ dimers, trimers, etc., which
are always present in a stable vapour, can in principle be determined by an exact statistical-mechanical
approach based on the cluster expansions of Mayer \cite{Mayer37}, Born
and Fuchs \cite{BF38}. As mentioned above, the modern molecular equations of state
from the SAFT \cite{CGJR90} and BACKONE \cite{MWF96} families are based
on this approach. With
some effort (which would involve developing a suitable concept of association), a molecular
equation of state could possibly be employed to compute quantities such as the monomer fraction as well
as higher-order cluster properties. In the literature,
it has already been attempted to extrapolate from the dimer fraction in a
stable vapour, obtained from the second virial coefficient, to the number of larger liquid nuclei
formed in a supersaturated vapour \cite{DM90, LFK94}.

While it is relatively uncommon to extrapolate from small clusters to larger ones,
an obverse approach which extrapolates from small (or zero) to high
curvature, is very widespread. The characteristic length
scale for the dependence of the surface tension on the radius is the Tolman length
\begin{equation}
   \tolman = \equimolar - \laplace,
\label{eqn:tolmanlength}
\end{equation}
introduced by Tolman \cite{Tolman48, Tolman49b} who applied the theoretical framework
of Gibbs \cite{Gibbs78a} to the adsorption $\adsorption$, i.e.\ the excess density,
at the spherical surface corresponding to the Laplace radius $\laplace$.
The Tolman length expresses the deviation of the equimolar
radius $\equimolar$, which corresponds to the spherical dividing surface with zero
adsorption, from the Laplace radius $\laplace$. It determines the dependence of the
surface tension on curvature according to the Tolman equation
\begin{equation}
   \frac{\differential\ln\surfacetension}{\differential\ln\laplace}
      = 1 + \frac{1}{2}\left(\frac{\tolman}{\laplace} + \left[\frac{\tolman}{\laplace}\right]^2
         + \frac{1}{3}\left[\frac{\tolman}{\laplace}\right]^3 \right)^{-1}.
\label{eqn:tolmaneq}
\end{equation}
Although Tolman \cite{Tolman49b} conjectured $\tolman$ to be positive and its
dependence on the radius to be of secondary importance, Eq.\ (\ref{eqn:tolmaneq})
is valid for any magnitude and dependence on $\laplace$ of the Tolman
length. However, its common interpretation as an expansion in terms of $1 \slash \radius$, i.e.\
\begin{equation}
   \frac{\surfacetension}{\planar\surfacetension}
      = \frac{1}{1 + 2\planar\tolman\laplace^{-1} + \dots},
\label{eqn:tolmanexp}
\end{equation}
has more recently come under criticism for a variety of reasons \cite{SVWZB09, TB11}, discussed
here in Sections \ref{sec:discussion} and \ref{sec:conclusion}. In any case, Eq.\ (\ref{eqn:tolmanexp}) has
the advantage of being based directly on the Tolman length $\planar\tolman$
and the surface tension $\planar\surfacetension$ of the planar vapour-liquid interface
which can be investigated experimentally in a stable state, as opposed to
nano-dispersed phases where this is in most cases practically impossible.

The Laplace radius $\laplace$ has the disadvantage of being defined by the
surface tension of the curved interface, which is thermodynamically well-defined,
but hard to determine. In consequence, it is often impossible to tell how
many molecules are inside a bubble or a droplet with the Laplace radius $\laplace$ (which would be
precisely known if an equimolar radius was specified), or which chemical
potential and pressure difference correspond to a particular value of $\laplace$.
Hence, considering that the dependence of the surface tension on curvature is
under dispute at present, Eq.\ (\ref{eqn:laplace}) contains two
unknowns and the Laplace radius is ill-defined at first.

For this reason, direct routes to the Tolman length have been
proposed which effectively eliminate the Laplace radius \cite{NBWBL91, FM09, GB09, LD10}.
The approach of Nij\-meijer et al.\ \cite{NBWBL91} as well as van Giessen
and Blokhuis \cite{GB09} can be formulated in terms of the equimolar surface tension, defined here by
\begin{equation}
   \equimolartension = \frac{\equimolar(\liq\pressure - \vap\pressure)}{2} = \frac{\surfacetension \equimolar}{\laplace},
\label{eqn:equimolartension}
\end{equation}
and its relation to the equimolar curvature $1 \slash \equimolar$.
In the planar limit, i.e.\ $1\slash\equimolar \to 0$,
the equimolar surface tension approaches the
surface tension of the planar vapour-liquid interface
\begin{equation}
   \lim_{1\slash\equimolar \to 0} \equimolartension
      = \left(\lim_{1\slash\equimolar \to 0} \surfacetension\right)
         \cdot \left(\lim_{1\slash\equimolar \to 0} \frac{\equimolar}{\laplace}\right) = \planar\surfacetension.
\end{equation}
An analogous relation holds for the derivative of the
surface tension with respect to curvature \cite{NBWBL91, HHSAEVMJ12}
\begin{eqnarray}
   \lim_{1\slash\equimolar \to 0} \left(\frac{\partial\equimolartension}{\partial(1\slash\equimolar)}\right)_\temperature
      & = & \lim_{1\slash\equimolar \to 0} \left(\frac{\partial\surfacetension}{\partial(1\slash\laplace)}\right)_\temperature \nonumber \\
         & = & - \planar\tolman\planar\surfacetension,
\end{eqnarray}
relating it to the Tolman length in the planar limit.

If the surface tension of the planar interface, rather than the actual surface tension of
the curved interface, is inserted into the Laplace equation
\begin{equation}
   \Delta\pressure = \liq\pressure - \vap\pressure = \frac{2\planar\surfacetension}{\capillarity},
\label{eqn:laplacecap}
\end{equation}
a direct route to $\tolman$ can be also be expressed in terms of the capillarity radius $\capillarity$,
defined by Eq.\ (\ref{eqn:laplacecap}). In this reformulation of Tolman's theory,
Eqs.\ (\ref{eqn:tolmanlength}) -- (\ref{eqn:tolmanexp}) transform to \cite{HHSAEVMJ12}
\begin{eqnarray}
   \eer & = & \equimolar - \capillarity, \label{eqn:defeer} \\
   \frac{\differential \ln \surfacetension}{\differential \ln (\planar\surfacetension\slash\capillarity)}
      & = & \frac{2}{3} \left(
         1 - \left[\frac{\planar\surfacetension(1 + \eer\invcapillarity)}{\surfacetension}\right]^3
            \right), \\
   \frac{\surfacetension}{\planar\surfacetension} & = & 1 + 2\frac{\planar\eer}{\capillarity} - 2\left(\frac{\planar\eer}{\capillarity}\right)^2 + \dots,
\end{eqnarray}
wherein $\eer$ is referred to as the excess equimolar radius. It
should be noted that in the planar limit, the Tolman length
and the excess equimolar radius are of the same magnitude, but
of opposite sign \cite{HHSAEVMJ12}
\begin{equation}
   \planar\tolman = -\planar\eer,
\end{equation}
despite their similar definition.
Here, this approach is applied to MD simulation results
for a box containing a gas bubble surrounded by a metastable liquid phase, cf.\ Section
\ref{subsec:results-b}, whereas in previous work employing the
same method \cite{HHSAEVMJ12}, only the case of a liquid droplet
surrounded by gas has been considered.

\section{Molecular simulation methodology}
\label{sec:method}

\subsection{Simulation software and molecular model}
\label{subsec:method-ls1}

\noindent
The present work applies MD simulation to the problems outlined above.
For this purpose, we employed the program
\textit{ls1 mardyn} \cite{NHBBEHGHBV13}, i.e.\ \qq{\textbf{l}arge \textbf{s}ystems
\textbf{1}st by \textbf{m}olecul\textbf{ar dyn}amics}. Eckhardt et al.\ \cite{EHBBHHKVHHBGNBB13} have
recently proven that \textit{ls1 mardyn} scales well in its
parallelized mode, delivering an almost
ideal speedup on modern supercomputer archi\-tectures and
even achieving a world record in system size for molecular simulation, with
$\absnum > 4 \times 10^{12}$. The scena\-rios considered here are
smaller by far, but partly require a long simulation time, so that an
efficient simulation code was a prerequisite for carrying out the
present study as well.

Since the theoretical state of the art leaves many qualitative
problems open for an investigation on the molecular level, the
Lennard-Jones truncated-shifted (LJTS) pair potential was selected as
the molecular model under consideration here. In reduced units, i.e.\
setting the Lennard-Jones size and energy
parameters $\LJsize = 1$ and $\LJenergy = 1$ (as well as the Boltzmann
constant $\kboltz = 1$) to unity, it is given by
\begin{equation}
   \potential(\distance) = \left\{
      \begin{array}{ll}
         4 \left[\left(\distance^{-12} - \distance^{-6}\right)
            - \left(\cutoff^{-12} - \cutoff^{-6}\right)\right],
               & \distance < \cutoff, \\
         0, & \distance \geq \cutoff, 
      \end{array}
   \right.
\label{eqn:LJTS}
\end{equation}
where $\distance$ is the distance between two molecules and
$\cutoff = 2.5$ is the cutoff radius.
Since the LJTS pair potential is a quantitatively precise model for
methane and several noble gases, including their vapour-liquid surface
tension \cite{VKFH06}, the present results also can be given a
realistic interpretation.

This choice of molecular model was also driven by the fact that
vapour-liquid interfacial properties of the LJTS fluid 
have been addressed in previous work from
several groups \cite{TGWCR84, NBWBL91, VKFH06, GB09, DB11b}, employing different
methods which can thus be compared directly.
The truncated-shifted cutoff, cf.\ Eq.\ (\ref{eqn:LJTS}), is
continuous in terms of the potential, but not with respect to the force
which has a discontinuity at $\distance = \cutoff$. The intermolecular
interaction is thereby strictly limited to radii smaller than
$\cutoff$, avoiding the complex issue of long-range cutoff corrections
in inhomogeneous systems \cite{Janecek06, WLHH13, YB13}.

\subsection{Influence of curvature on vapour-liquid equilibria}
\label{subsec:method-ab}

\noindent
Extending previous work on the excess equimolar radius of liquid
droplets \cite{HHSAEVMJ12}, a series of MD simulations was conducted
for volumes containing a LJTS gas bubble in equilibrium with a subsaturated liquid.
The simulations were carried out in the canonical ensemble with a
periodic boundary condition.
The initial conditions were chosen such that one single bubble existed
in the centre of the simulation box. The size of that bubble was
controlled by choosing the number of molecules and the
simulation volume appropriately.
As pointed out by Fisher and
Wortis \cite{FW84} as well as Reguera et al.\ \cite{RBDR03}, such equilibria can be
thermodynamically stable, even if the phase (here, the liquid) which
surrounds the dispersed phase (here, the gas bubble) would be
metastable in a corresponding homo\-geneous state.
Obviously, they can only be thermodynamically stable when the simulation
volume is relatively small -- the precise conditions depend on the
equation of state of the fluid -- and for configurations containing a
single gas bubble.

The present MD simulations are therefore concerned with the
scenario where a single gas bubble is surrounded by a subsaturated
liquid phase, under equilibrium conditions for the pure LJTS fluid.
The temperature was specified to be $\temperature = 0.75$, i.e.\ about
$70$ \% of the critical temperature \cite{VKFH06}, and controlled by a
velocity rescaling thermostat. The number of mole\-cules $\absnum$ and
the simulation volume $\volume$ were varied as indicated in Table
\ref{tab:bedhotiya}. An equilibration was conducted for at least
$400$ $000$ time steps, with an integration time step of $0.003$ in
reduced units. 
A novel shading approach for the visualization of point-based
datasets, which makes it easier to analyze the morphology of an
interface on the molecular level \cite{ESH13}, was applied to
individual configurations, cf.\ Fig.\ \ref{fig:pointao}.
Density profiles were determined by binning over
several averaging intervals of at least $200$ $000$ time steps until
the profiles of were found to converge. The system of coordinates
was shifted continuously, following the random motion of the bubble
to keep its centre in the origin.

From these density profiles, cf.\ Fig.\ \ref{fig:densityprofiles},
the equimolar radius $\equimolar$, the capillarity radius
$\capillarity$, and thus the excess equimolar
radius $\eer = \equimolar - \capillarity$ were determined by following
theoretical approach discussed in Section \ref{subsec:theory-bd}.
However, in contrast with the method previously established for the
simulation of liquid drops \cite{HHSAEVMJ12}, the pressure
$\vap\pressure$ inside the gas bubble, and thereby the capillarity
radius
\begin{equation}
   \capillarity = \frac{2\planar\surfacetension}{\liq\pressure - \vap\pressure},
\label{eqn:detcap}
\end{equation}
was not determined here from the density profile on the vapour side.
Instead, only the density $\liq\density$ of the subsaturated liquid surrounding
the bubble was extracted from the density profile by extrapolating to
infinite distance from the centre of the bubble. The liquid phase can
very accurately be sampled within the MD simulation and is much closer
to bulk-like behaviour than the vapour phase here.

It should be recalled that the values
of $\liq\pressure$ and $\vap\pressure$ which the theory requires are
not the actual mechanical pressures outside and inside, but those of
the respective subsaturated bulk phases at the same chemical potential (cf.\ the
discussion in Section \ref{subsec:theory-bd}). Therefore, the pressure of the
vapour phase was determined here, accordingly, from the thermal and chemical
equilibrium condition by means of an emprical fifth order virial
equation of state \cite{HMVGNBMJ12}.
For the subsequent discussion, however, this methodical issue is of
minor importance, since $\Delta\pressure$ is dominated by the liquid
term, which was obtained here by the same extrapolation method as
previously published \cite{HHSAEVMJ12}.

In a second series of simulations, the qualitative influence of
curvature was considered. For this purpose, canonical ensemble MD
simulations were carried out for a bubble (surrounded by a
subsaturated liquid), a droplet (surrounded by a supersaturated
vapour), and a system consisting of a vapour and a liquid slab
separated by planar interfaces.
For these systems, the chemical
potential was computed by applying the Widom test particle
method \cite{Widom82} with $\absnum$ test insertions and
deletions every $16$ time steps, where $\absnum$ is the number of
particles in the system. To compensate for the additional
computational effort, the averaging interval for constructing the
profiles was reduced to $10$ $000$ time steps here.

The simulation conditions were chosen
here such that the radii of the droplet and the bubble were about
$8.5$, while the thickness of the vapour and the liquid slab was about
$12.5$, complementing previous simulation results \cite{HBCDFRWVH13}.
The subsaturation (for bubbles) or supersaturation (for droplets) was
determined from the deviation
\begin{equation}
   \Delta\chempot = \chempot - \chempot_\mathrm{sat}
\end{equation}
between the chemical potential in the system with the curved interface
and the value $\chempot_\mathrm{sat}$ computed at the planar
interface. On this basis, $\liq\pressure$ as well as $\vap\pressure$
for the second series of simulations were calculated from the virial
equation for the LJTS fluid \cite{HMVGNBMJ12}.

\section{Simulation results}
\label{sec:results}
\label{subsec:results-b}

\noindent
The density profiles of gas bubbles in equilibrium with subsaturated
liquid phases, which were obtained by MD simulation in the canonical
ensemble, are shown in Fig.\ \ref{fig:densityprofiles}. The density
in the centre of the bubble should be expected to approach the
saturated vapour density, i.e.\ $\vap\density(\temperature = 0.75) =
0.0124$ \cite{VKFH06}, in the limit of an infinitely large bubble
($\laplace \to -\infty$), which corresponds to the transition to a
planar interface.
The present simulation results confirm this, cf.\ Tab.\
\ref{tab:bedhotiya} and the results for $\equimolar = -28$ shown
therein. Moreover, deviations of the vapour density from its value at
saturation over a planar interface $\density''_\mathrm{sat}$ are
observed for small bubbles, cf.\ Fig.\ \ref{fig:density-deviation}.
This deviation is caused by two qualitatively distinct effects:
\begin{enumerate}
   \item{} For relatively large
           bubbles ($-\infty < \equimolar < -9$), the density in the
           centre decreases as the size of the bubble becomes smaller.
           The minimal gas density observed in the
           present series of simulations, which is significantly
           below $0.01$, is found in the centre of the bubble
           with $\equimolar = -8.7$.
   \item{} For even smaller bubbles ($-9 < \equimolar < 0$), the
           density in the centre increases again. In the smallest case
           considered here, i.e.\ $\equimolar = -5.6$, the gas phase
           is found to be much denser than that which coexists with
           the liquid at a planar interface, cf.\ Fig.\ \ref{fig:densityprofiles}.
\end{enumerate}
In Tab.\ \ref{tab:bedhotiya}, numerical results are shown that were
obtained from these simulations by following the approach outlined
in Section \ref{subsec:method-ab}, based on liquid densities
extracted from the present density profiles.
The density of the liquid phase surrounding the gas bubble was found
to be subsaturated in all cases. In particular, as shown in Fig.\
\ref{fig:liq-density}, smaller bubbles consistently correspond to
smaller liquid densities here, in agreement with capillary theory.

The excess equimolar radius $\eer$ was found
to be positive in all cases, indicating a deviation from the capillarity
approximation where, to first order in $1\slash\radius$, the surface tension of a droplet is
larger and the surface tension of a bubble is smaller than that of the
planar vapour-liquid interface.

Results for the chemical potential of bubbles, planar slabs, and
droplets, cf.\ Tab.\ \ref{tab:interpol}, corroborate the thermodynamic
approach to the analysis of curved interfaces outlined in Section
\ref{subsec:theory-bd}.
The chemical potential of droplets (and the vapour surrounding them)
was consistently found to be higher than the value at saturation over
a planar interface. Obversely, nanoscopic gas bubbles and the liquid
phase surrounding them are subsaturated, and the deviation from
$\chempot_\mathrm{sat}$ increases as the dispersed phase becomes
smaller. 

\section{Discussion}
\label{sec:discussion}

\noindent
As pointed out above, it is
one of the observations from the present simulations of curved
vapour-liquid interfaces that a nanobubble with a diameter larger than $5$ nm, roughly
corresponding to $\abs{\radius} > 6$ for the LJTS fluid \cite{VKFH06}, has a smaller
density than the bulk vapour at the dew
line (see Fig.\ \ref{fig:densityprofiles}). This is the behaviour
which should be expected from capillary theory, based on Gibbs'
thermodynamic interpretation of the Laplace equation. It was also
confirmed that the subsaturated density corresponds to a subsaturated
chemical potential ($\chempot < \chempot_\mathrm{sat}$), cf.\ Tab.\
\ref{tab:interpol}, in agreement with the thermodynamic discussion
of the curvature influence on fluid phase coexistence (see Fig.\
\ref{fig:mu-p-diagram}).

On the other hand, the vapour density in the centre of the bubble
was found to increase again for even smaller bubbles,
eventually exceeding the dew density. This is not paralleled by an
increase, but rather by a further decrease of the liquid density,
cf.\ Fig.\ \ref{fig:liq-density}, which suggests that
in terms of the chemical potential, these extremely small bubbles
are subsaturated as well. This implies that among the two effects
present for the gas density, only one affects the surrounding liquid
as well, suggesting the following interpretation: Both phases, vapour
and liquid, tend to become subsaturated due to \textit{interfacial
curvature}, cf.\ Fig.\ \ref{fig:mu-p-diagram}.
The density in the centre of the bubble, however,
experiences an additional obverse influence due to a size-dependent
phenomenon which is distinct from curvature.

The density profiles, cf.\ Fig.\ \ref{fig:densityprofiles}, suggest
that the density of the gas phase
is increased \textit{not due to curvature}, which tends to reduce
$\chempot$ and thereby also $\vap\density$, but because there is
not enough space available in radial direction for the density profile
to converge to the bulk density that would correspond to the
respective value of $\chempot$. Therefore,
this second effect should be ascribed to the
extremely \textit{small diameter} of the nanobubbles.
In the present simulations, however, no analogous effect is found in
the liquid phase. This may be related to the fact that the liquid has a much higher
density, so that a perturbation which is significant for $\vap\density$ may well appear
to be negligible in comparison with $\liq\density$.

This parallels the recent discovery, by Malijevsk\'y and
Jackson \cite{MJ12}, of two distinct size-dependent effects concerning
the surface tension of nanodroplets: The Tolman length $\tolman$ was
found to be negative, causing the surface tension to increase over
its planar value. The leading term, which dominates this effect for
relatively large radii, is proportional to $1\slash\radius$.
Extremely small droplets, however, exhibit a reduced surface tension.
From an empirical correlation, Malijevsk\'y and Jackson \cite{MJ12} found this
contribution to $\surfacetension$, which acts obversely to Tolman's curvature effect, to be
proportional to $1\slash\radius^3$.

In a subsequent study of Werth et al.\ \cite{WLHH13}, the surface tension of thin
planar liquid slabs with a thickness of $\thickness$ was found to be
reduced, with respect to the macroscopic vapour-liquid surface
tension, by a term proportional to $\textnormal{$1\slash\thickness^3$}$. Furthermore,
density profiles revealed the density in the centre of these nanoslabs
to deviate from the density of the saturated bulk liquid by a term
proportional to $\textnormal{$1\slash\thickness^3$}$ as well, suggesting that the
two phenomena are related expressions of a single effect which
is caused by the small thickness of the interface \cite{WLHH13}.

The present results complement the picture by proving that for gas
bubbles, distinct effects due to curvature and due to the small diameter,
respectively, can be detected as well, cf.\ Fig.\ \ref{fig:density-deviation}. Furthermore, the excess
equimolar radius was found to be positive here, corresponding to a
negative Tolman length, which confirms the tendency found by
Malijevsk\'y and Jackson \cite{MJ12}. For the surface tension of a
bubble, however, these two effects do not counteract but rather
reinforce each other, since both the curvature effect from the Tolman
equation (with $\tolman < 0$ and a negative curvature) and the
small-diameter effect contribute to a reduction of $\surfacetension$.

This is confirmed by an analysis
following the approach of Nijmeijer et al.\ \cite{NBWBL91}
as well as van Giessen and Blokhuis \cite{GB09}, applied to the
previous simulations of single droplets \cite{HHSAEVMJ12} and the present
simulations of single bubbles, cf.\ Tab.\ \ref{tab:nijmeijer}. In particular,
the equimolar surface tension $\equimolartension$,
cf.\ Eq.\ (\ref{eqn:equimolartension}), is consistently smaller for a
gas bubble than for a liquid droplet.

The surface tension of the
planar vapour-liquid interface of the LJTS fluid at
$\temperature = 0.75$, which is $\planar\surfacetension = 0.493$ according
to the correlation of Vrabec et al.\ \cite{VKFH06}, deviates relatively
little from the $\equimolartension$ values found for the droplet. The
equimolar surface tension of bubbles from the present simulations,
however, is significantly smaller than $\planar\surfacetension$. This
is also consistent with the previous result that $\tolman$ and $\eer$
are relatively small for a droplet \cite{HHSAEVMJ12}, whereas for a
bubble, relatively large positive values of $\eer$ were obtained
here, cf.\ Tab.\ \ref{tab:bedhotiya}, corresponding to a negative
Tolman length $\tolman$.

On the basis of Hadwiger's theorem \cite{Hadwiger57}, it has been argued that
the influence of geometry on the surface tension needs to be proportional
to the mean curvature, the Gaussian curvature, or linear combinations
thereof \cite{KRM04}.
Such an interpretation of Hadwiger's theorem would explicitly rule
out any curvature-independent effect.
This cannot be upheld in the light of the present discussion,
since the small-diameter effect, which has
now been detected for bubbles as well as for droplets, exists analogously for
planar slabs where curvature is strictly absent \cite{WLHH13}.

Beside the curvature and the diameter,
further aspects of confinement may significantly influence
vapour-liquid coexistence in small systems.
In the past, such effects have largely been discussed
separately from each other. A unified approach to describing the
thermophysical properties of nano-dispersed fluid phases would have to
account for various size-dependent phenomena in a consistent way:
\begin{itemize}
   \item{} The effect of curvature, cf.\ Tolman \cite{Tolman49b} and the
           pre\-sent discussion.
   \item{} The effect of a small diameter, cf.\ Werth et al.\ \cite{WLHH13}
           and the present discussion.
   \item{} The effect of the capillary wave cutoff, cf.\ Sengers and van
           Leeuwen \cite{SL89}. The small circumference of the
           nano-dispersed phase imposes a restriction on the available
           modes, each of which contributes to the interfacial free
           energy.
   \item{} The effect of fluctuations, cf.\ Reguera et al.\ \cite{RBDR03}.
           For a small dispersed phase, which is surrounded by a large bulk
           phase, the temperature, the density, and the volume can fluctuate
           significantly.
\end{itemize}
A theoretical approach which accounts for the interplay between these
phenomena and yet retains the simplicity of Tolman's equation or the
inverse cube law for the diameter effect is missing so far, however.
Consequently, where no experimental data are available, molecular
simulation is at present the only viable method for predicting the
properties of nano-dispersed phases.








\section{Conclusion}
\label{sec:conclusion}

\noindent
Molecular simulation is feasible up to the micrometre length
scale by massive\-ly-paral\-lel molecular dynamics today, facilitating an
analysis of the size dependence for interfacial phenomena which it
would otherwise be relatively hard to investigate in a
reliable way.
By molecular simulation, which is firmly founded on
statistical mechanics, such effects can be rigorously investigated. 
In combination with the previous research of Malijevsky and Jackson
\cite{MJ12} on droplets as well as Werth et al.\ \cite{WLHH13} on
thin slabs, present results on gas bubbles complete the recent body
of work on the interplay of distinct effects due to a high curvature
of the interface and a small diameter of the dispersed phase, respectively.

Regarding the thermodynamic properties of nano-dispersed fluid phases,
Tr\"oster and Binder \cite{TB11} have recently pointed out that
as for small droplets there is, for instance, a significant
deviation from the planar surface tension, but this
effect does not consistently agree with
the Tolman equation, \qq{neither the capillarity approximation nor the
Tolman parametrization [$\dots$] should be employed in any serious
quantitative work.}
The present analysis supports this conclusion.
Instead of the Tolman equation, a new theoretical framework needs to
be developed to describe the various size-dependent effects related
to the curvature, the diameter, and possibly the circumference as
well as the volume, which controls the magnitude of fluctuations,
in a coherent way.

\smallskip

\textit{Acknowledgment.\/}
The authors would like to thank BMBF for funding the SkaSim project,
DFG for funding the Collaborative Research Centre MICOS (SFB 926),
Akshay Bedhotiya for carrying out some of the molecular simulations of gas bubbles,
Sebastian Eichelbaum and Mario Hlawitschka as well as Gerik Scheuermann for employing one of the present gas bubbles as a test case for PointAO shading,
Kai Sundmacher for his encouragement,
and Jadran Vrabec for his continuous support,
as well as Stefan Becker, Ruslan Davidchack, Sergey Lishchuk, Andrew Masters, Erich M\"uller, and Stephan Werth for fruitful discussions.
The present work was conducted unter the auspices of the Boltz\-mann-Zuse Society of Computational Mo\-le\-cu\-lar Engineering (BZS),
and the MD simulations were carried out on the \textit{elwetritsch} cluster, Regionales Hoch\-schul\-rechen\-zentrum Kai\-sers\-lau\-tern, within the scientific computing project TUKL-MSWS.


\newpage

\begin{table}[t!]
\centering
\caption{}
\label{tab:bedhotiya}
\end{table}

\begin{center}
\begin{tabular}{cc||cc||cccc|cr} \hline
$\absnum$ & $\volume$ & $\density'_\infty$ & $\density''_0$ & $\liq\pressure(\density'_\infty)$ & $\vap\pressure(\density'_\infty)$ & $- \equimolar$ & $- \capillarity$ & $\eer$ \\ \hline

$\phantom{00}7$ $303$ & $\phantom{0}10$ $648$ & $0.7360$ & $0.023$ & $-0.16\phantom{0}$ & $0.0061$ & $\phantom{0}5.6$ & $\phantom{1}6.1$ & $0.5$ & \\
$\phantom{00}9$ $551$ & $\phantom{0}13$ $824$ & $0.7365$ & $0.015$ & $-0.15\phantom{0}$ & $0.0061$ & $\phantom{0}5.9$ & $\phantom{1}6.2$ & $0.3$ & \\
$\phantom{0}18$ $107$ & $\phantom{0}27$ $000$ & $0.7457$ & $0.008$ & $-0.093$ & $0.0069$ & $\phantom{0}8.7$ & $\phantom{0}9.9$ & $1.2$ & \\
$\phantom{0}42$ $474$ & $\phantom{0}64$ $000$ & $0.7493$ & $0.010$ & $-0.068$ & $0.0072$ & $12.1$ & $13.1$ & $1.0$ & \\
$\phantom{0}34$ $944$ & $\phantom{0}54$ $872$ & $0.7508$ & $0.009$ & $-0.058$ & $0.0074$ & $12.6$ & $15.1$ & $2.5$ & \cite{HBCDFRWVH13} \\
$\phantom{0}75$ $794$ & $117$ $649$ & $0.7521$ & $0.011$ & $-0.048$ & $0.0075$ & $16.0$ & $17.7$ & $1.7$ & \\
$122$ $232$ & $195$ $112$ & $0.7538$ & $0.011$ & $-0.035$ & $0.0077$ & $20.0$ & $23.0$ & $3.0$ & \\
$263$ $163$ & $438$ $976$ & $0.7556$ & $0.012$ & $-0.022$ & $0.0079$ & $28.0$ & $32.8$ & $4.8$ & \\ \hline
\end{tabular}
\end{center}

\bigskip

   \noindent
   Number of particles $\absnum$ and simulation volume $\volume$ for
   a series of canonical ensemble MD simulations of LJTS bubbles in
   equilibrium (at $\temperature = 0.75$). The density
   $\density'_\infty$ of the liquid phase was obtained
   here by extrapolating the density profiles to infinite distance
   from the centre of the bubble. It is subsaturated with respect to
   the bubble density $\density'_\mathrm{sat}(\temperature) = 0.759$ of the bulk fluid. 
   The gas density $\density''_0$ was determined
   in a region closer than $1.5$ to the centre of the bubble.
   The thermodynamic liquid and vapour pressures $\liq\pressure(\density'_\infty)$
   and $\vap\pressure(\density'_\infty)$ to be used within the Gibbs approach,
   respectively, were computed from the
   subsaturated liquid density by a fifth-order virial expansion \cite{HMVGNBMJ12};
   they may deviate from the eigenvalues of the mechanical pressure tensor outside
   and inside the bubble.
   From the equimolar and capillarity radii $\equimolar$ and
   $\capillarity$, respectively, which are negative by the convention
   employed here, the excess equimolar radius $\eer = \equimolar - \capillarity$ was obtained.

\newpage

\begin{table}[t!]
\centering
\caption{}
\label{tab:interpol}
\end{table}

\begin{center}
\begin{tabular}{cc||cc||ccr} \hline
$\absnum$ & $\volume$ & $\equimolar$ & $\chempot$ & $\liq\pressure(\Delta\chempot)$ & \\ \hline

$18$ $107$ & $27$ $000$ & $-\phantom{0}8.7$ & $-3.55(3)$ & $-0.13(4)\phantom{00}$ & \\
$34$ $944$ & $54$ $872$ & $-12.6$ & $-3.51(2)$ & $-0.10(3)\phantom{00}$ & \cite{HBCDFRWVH13} \\ \hline

$\absnum$ & $\volume$ & $\thickness_\density$ & $\chempot_\mathrm{sat}$ & $\pressure_\mathrm{sat}$ & \\ \hline

$\phantom{0}7$ $079$ & $18$ $341$ & $\phantom{-0}8.5$ & $-3.37(2)$ & $\phantom{-}0.0084\phantom{(0)}$ & \cite{HBCDFRWVH13} \\
$10$ $409$ & $26$ $971$ & $\phantom{-}12.5$ & $-3.37(2)$ & $\phantom{-}0.0084\phantom{(0)}$ & \\ \hline

$\absnum$ & $\volume$ & $\equimolar$ & $\chempot$ & $\liq\pressure(\Delta\chempot)$ & \\ \hline

$\phantom{0}2$ $425$ & $27$ $000$ & $\phantom{-0}8.6$ & $-3.28(6)$ & $\phantom{-}0.08(6)\phantom{00}$ & \\
$\phantom{0}6$ $844$ & $54$ $872$ & $\phantom{-}12.4$ & $-3.31(4)$ & $\phantom{-}0.05(5)\phantom{00}$ & \cite{HBCDFRWVH13} \\ \hline
\end{tabular}
\end{center}

\bigskip

   \noindent
   Results for bubbles (top), planar slabs (middle),
   and droplets (bottom) from equilibrium MD
   simulation of the LJTS fluid in the canonical ensemble with $\absnum$ particles and a
   simulation volume of $\volume$ at a temperature of $\temperature =
   0.75$, where the equimolar radii $\equimolar$ and slab thicknesses
   $\thickness_\density$ were determined from density profiles,
   while the chemical potential $\chempot$ was computed by
   Widom's test particle method \cite{Widom82}. The liquid
   pressure $\liq\pressure$ was calculated from the
   deviation $\Delta\chempot$ between the chemical potential at the
   planar and curved interfaces on the basis of an equation of state
   \cite{HMVGNBMJ12}. The error for $\chempot$ and $\liq\pressure$,
   respectively, is indicated in parentheses, where the error of
   is of the same magnitude as the last given digit.

\newpage

\begin{table}[t!]
\centering
\caption{}
   \label{tab:nijmeijer}
\end{table}

\begin{center}
\begin{tabular}{cc||cc||c} \hline
$\absnum$ & $\volume$ & $\liq\pressure - \vap\pressure$ & $1\slash\equimolar$ & $\equimolartension$ \\ \hline

   $\phantom{00}7$ $303$ & $\phantom{0}10$ $648$ & $-0.16(1)\phantom{0}$ & $-0.180\phantom{0}$ & $0.45(4)$ \\
   $\phantom{00}9$ $551$ & $\phantom{0}13$ $824$ & $-0.159(5)$ & $-0.169\phantom{0}$ & $0.47(2)$ \\
   $\phantom{0}42$ $474$ & $\phantom{0}64$ $000$ & $-0.075(6)$ & $-0.0827$ & $0.46(3)$ \\
   $\phantom{0}75$ $794$ & $117$ $649$ & $-0.056(7)$ & $-0.0626$ & $0.44(5)$ \\
   $122$ $232$ & $195$ $112$ & $-0.043(4)$ & $-0.0500$ & $0.43(4)$ \\ \hline

$\absnum$ & $\volume$ & $\liq\pressure - \vap\pressure$ & $1\slash\equimolar$ & $\equimolartension$ \\ \hline

   $\phantom{0}15$ $237$ & $166$ $375$ & $\phantom{-}0.060(2)$ & $\phantom{-}0.0626$ & $0.48(2)$ \\
   $\phantom{0}12$ $651$ & $140$ $608$ & $\phantom{-}0.065(2)$ & $\phantom{-}0.0668$ & $0.49(2)$ \\
   $\phantom{0}10$ $241$ & $110$ $592$ & $\phantom{-}0.070(1)$ & $\phantom{-}0.0716$ & $0.49(1)$ \\
   $\phantom{00}6$ $619$ & $\phantom{0}74$ $088$ & $\phantom{-}0.080(2)$ & $\phantom{-}0.0831$ & $0.48(1)$ \\
   $\phantom{00}5$ $161$ & $\phantom{0}54$ $872$ & $\phantom{-}0.085(3)$ & $\phantom{-}0.0902$ & $0.47(1)$ \\
   $\phantom{00}3$ $762$ & $\phantom{0}39$ $304$ & $\phantom{-}0.102(2)$ & $\phantom{-}0.100\phantom{0}$ & $0.51(1)$ \\
   $\phantom{00}1$ $418$ & $\phantom{0}21$ $952$ & $\phantom{-}0.15(1)\phantom{0}$ & $\phantom{-}0.145\phantom{0}$ & $0.52(4)$ \\ \hline
\end{tabular}
\end{center}

\bigskip

   \noindent
   Number of molecules $\absnum$, simulation volume $\volume$,
   pressure difference $\liq\pressure - \vap\pressure$ between the
   coexisting fluid phases, equimolar curvature $1\slash\equimolar$,
   and equimolar surface tension $\equimolartension$,
   cf.\ Eq.\ (\ref{eqn:equimolartension}), from the present
   MD simulations of gas bubbles (top) as well as the MD
   simulations of liquid droplets (bottom) from previous
   work \cite{HHSAEVMJ12}, for the LJTS fluid
   at $\temperature = 0.75$.
   Numbers in parentheses represent the error, with the
   magnitude corresponding to that of the last given digit
   (only results for $\equimolartension$ with an error of
   $0.05$ or less are shown here).

\newpage

\begin{figure}[t!]
\centering
\caption{
}
\label{fig:kasten}
\end{figure}

\begin{center}
\includegraphics[width=7.5cm]{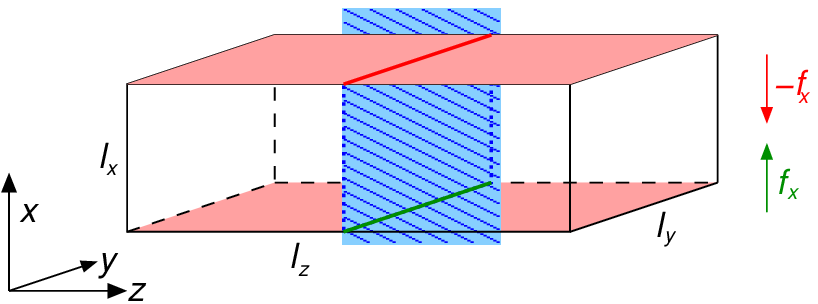}
\end{center}

\bigskip

   \noindent
   Diagram illustrating the mechanical definition of the surface
   tension. The two faces of the box with an orientation perpendicular
   to the $\coordx$ axis experience forces in opposite directions,
   expressing the tendency of an interface situated in the centre of
   the box to contract. The magnitude of the force $\force_\coordx$ is proportional to
   the surface tension $\surfacetension$ and the length of the contact line $\length_\coordy$.

\newpage

\begin{figure}[t!]
\centering
\caption{
}
\label{fig:mu-p-diagram}
\end{figure}

\begin{center}
\includegraphics[width=7.5cm]{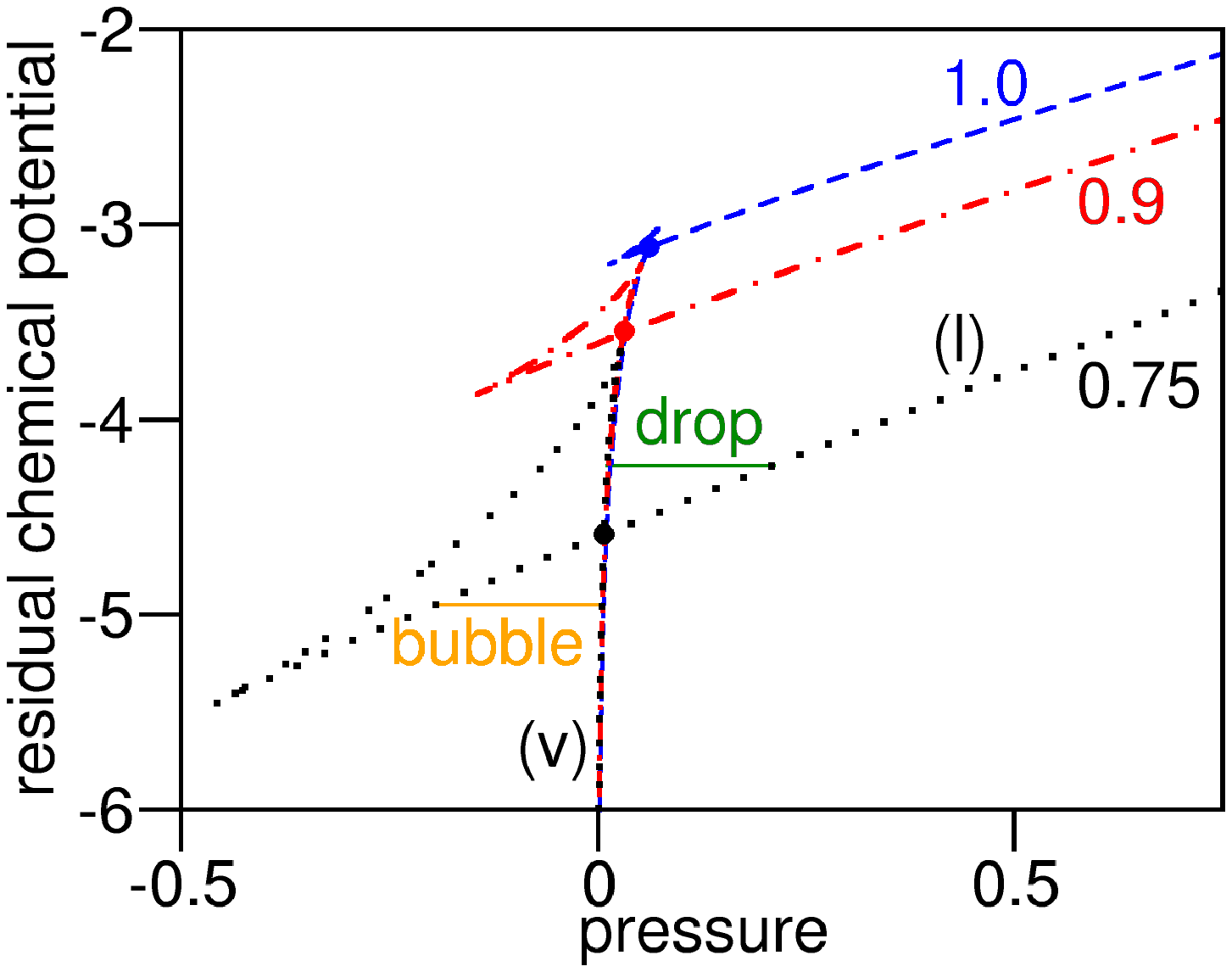}
\end{center}

\bigskip

   \noindent
   Isothermal dependence of the residual chemical potential $\chempotres$, cf.\ Eq.\
   (\ref{eqn:chempotres}), on the pressure $\pressure$ from a virial
   expansion \cite{HMVGNBMJ12} for the truncated-shifted Lennard-Jones
   potential at reduced temperatures of $0.75$ ($\cdots$), $0.9$
   ($\cdot$ -- $\cdot$), and $1.0$ (-- --). The plot extends over
   the whole range of vapour (v) and liquid (l) densities including stable, metastable and unstable
   states. Self-intersections of the isotherms ($\bullet$) correspond to
   the phase equilibrium condition at a planar interface, i.e.\
   $\liq\chempot = \vap\chempot = \chempot_\mathrm{sat}(\temperature)$
   and $\liq\pressure = \vap\pressure = \pressure_\mathrm{sat}(\temperature)$.
   Solid horizontal lines: Vapour-liquid equilibrium at a
   curved interface characterized by the Laplace equation, cf.\ Eq.\
   (\ref{eqn:laplace}), where the reduced temperature is $0.75$ and
   the pressure is smaller outside than for the dispersed phase,
   which is confined by the interface, with a pressure difference
   of $\liq\pressure - \vap\pressure = \pm$ $0.2$ in reduced units.

\newpage

\begin{figure}[t!]
\centering
\caption{
}
\label{fig:pointao}
\end{figure}

\begin{center}
\includegraphics[width=7.5cm]{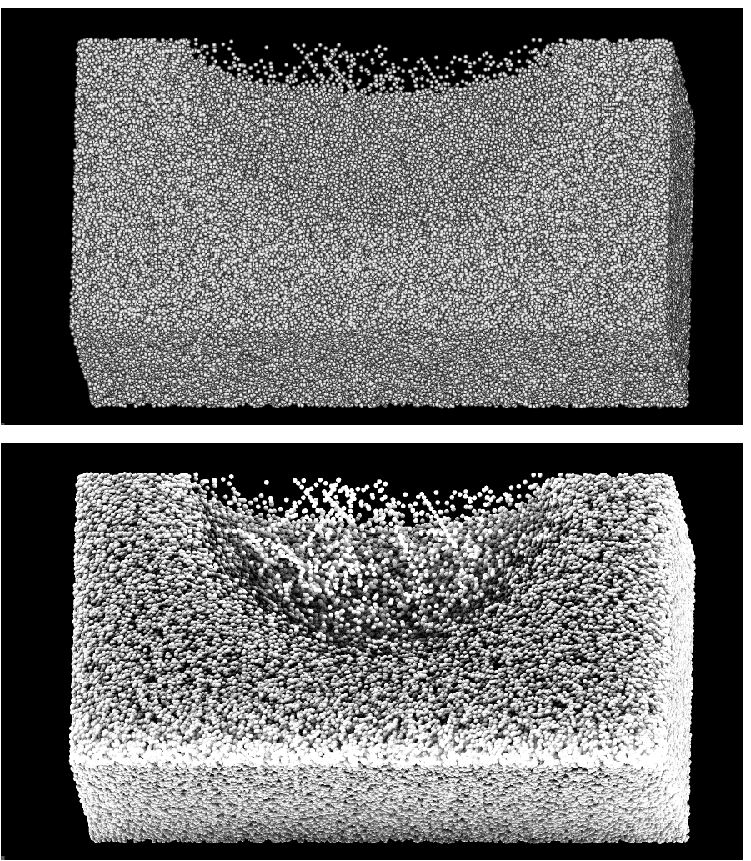}
\end{center}

\bigskip

   \noindent
   Visualization of the same configuration by Phong shading (top)
   as opposed to the novel PointAO shading
   algorithm (bottom), cf.\ Eichelbaum et al.\ \cite{ESH13}.

\newpage

\begin{figure}[t!]
\centering
\caption{
}
\label{fig:densityprofiles}
\end{figure}

\begin{center}
\includegraphics[width=7.5cm]{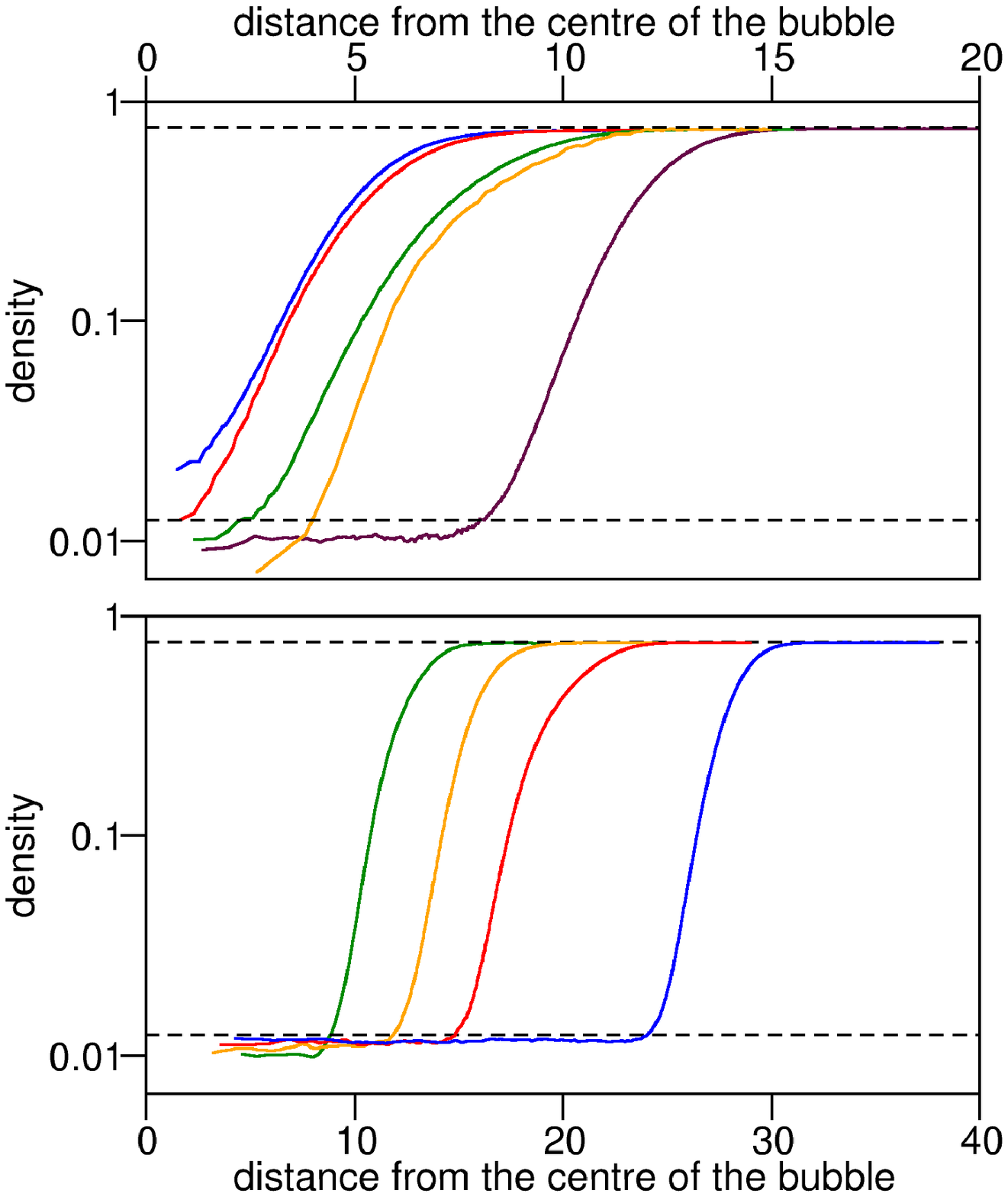}
\end{center}

\bigskip

   \noindent
   Density profiles of bubbles in equilibrium with a subsaturated
   liquid phase from MD simulation of the LJTS fluid in the canonical
   ensemble (---) in comparison with the vapour and liquid
   densities at saturation (-- --), for a temperature of $\temperature = 0.75$.
   Top: Results for five relatively small bubbles with equimolar
   radii $\equimolar = -5.6$, $-5.9$, $-8.0$, $-8.7$, and $-12.1$
   (from left to right); Bottom: Results for four relatively large
   bubbles with $\equimolar = -12.6$, $-16.0$, $-20.0$,
   and $-28.0$ (from left to right),
   including simulation results for $\equimolar = -12.6$ from
   previous work \cite{HBCDFRWVH13}.

\newpage

\begin{figure}[t!]
\centering
\caption{
}
\label{fig:density-deviation}
\end{figure}

\begin{center}
\includegraphics[width=7.5cm]{fig-density-deviation}
\end{center}

\bigskip

   \noindent
   Density in the centre over the equimolar radius of gas bubbles,
   which is negative here by convention, from present MD simulations of the LJTS fluid in
   the canonical ensemble at $\temperature = 0.75$ ($\circ$), including a data point for
   $\equimolar = -12.6$ from previous work \cite{HBCDFRWVH13}, in
   comparison with the vapour density at saturation (---) and a
   thermodynamic prediction from the capillarity
   approximation (-- --), considering curvature effects only and
   assuming $\surfacetension = \planar\surfacetension$ (and
   hence $\laplace = \capillarity = \equimolar$), as well as a correlation
   which also includes a deviation from the capillarity approximation
   proportional to the inverse cube of the radius ($\cdots$), i.e.\
   $\Delta\density = -1.5 \slash \radius_\density^{-3}$, due to the
   small-diameter effect found by Malijevsk\'y and
   Jackson \cite{MJ12}.

\newpage

\begin{figure}[t!]
\centering
\caption{
}
\label{fig:liq-density}
\end{figure}

\begin{center}
\includegraphics[width=7.5cm]{fig-liq-density}
\end{center}

\bigskip

   \noindent
   Liquid density $\density'_\infty$, obtained by extrapolating the
   density profiles from present MD simulations to an infinite distance
   from the centre of the
   gas bubble ($\circ$), over the equimolar radius  $\equimolar$, which is negative by the convention
   employed here, for the LJTS fluid in the canonical ensemble at
   $\temperature = 0.75$, including a data point for
   $\equimolar = -12.6$ from previous work \cite{HBCDFRWVH13}, in
   comparison with the liquid density at saturation (---) as well as a
   thermodynamic prediction from the capillarity
   approximation (-- --), considering curvature effects only and
   assuming $\surfacetension = \planar\surfacetension$ (and
   hence $\laplace = \capillarity = \equimolar$).

\end{document}